\documentclass[aps,prl,twocolumn,groupedaddress]{revtex4}
\usepackage{amsmath}
\usepackage{amssymb}
\usepackage{amsfonts}
\usepackage[dvips]{graphicx}
\usepackage[dvips, bookmarks, colorlinks=true]{hyperref}
\begin{document}

\title{Perfect dc conductance of a finite width Mott insulator sandwiched between metallic leads at zero temperature: a quantum emergent phenomenon in strongly correlated multilayers}

\author{H. Zenia$^1$, J. K. Freericks$^1$, H. R. Krishnamurthy$^{1,2,3}$, and Th. Pruschke$^4$}
\affiliation{$^1$Department of Physics, Georgetown University, Washington, DC 20057 USA\\
$^2$Centre for Condensed Matter Theory, Department of Physics, Indian Institute of Science,
 Bangalore 560012, India\\
$^3$Condensed Matter Theory Unit,
Jawaharlal Nehru Centre for Advanced Scientific Research,
Bangalore 560064, India\\
$^4$Institute for Theoretical Physics, University of G\"ottingen, Friedrich-Hund-Platz 1,
D-37077 G\"ottingen, Germany}

\begin{abstract}
Using inhomogeneous dynamical mean-field theory, we argue that the normal-metal proximity effect forces any finite number of ``barrier'' planes that are described by the (paramagnetic) Hubbard model and sandwiched between semi-infinite metallic leads to always be a Fermi liquid at $T=0$. This then implies that the inhomogeneous system restores lattice periodicity at zero frequency, has a well-defined Fermi surface, and should display perfect (ballistic) conductivity or ``transparency''. These results are, however, fragile with respect to finite $\omega$, $V$, $T$, disorder, or magnetism, all of which restore the expected quantum tunneling regime through a finite-width Mott insulator. Our formal results are complemented by numerical renormalization group studies on small thickness barriers that illustrate under what circumstances this behavior might be seen in real experimental systems.
\end{abstract}

\maketitle

In recent years, there has been significant activity in the study of multilayered heterostructures constructed out of strongly correlated materials. One of the reasons for interest in these systems is that new physical phenomena emerge in the heterostructures that are
absent in the bulk materials they are composed from, like the appearance of two-dimensional electron gases at the interfaces between band and Mott insulators~\cite{hwang} and their low-temperature superconductivity~\cite{thiel06}.  Here we show the possibility of another such striking, quantum emergent phenomenon, that {\em at sufficiently low temperatures, a finite width Mott insulator sandwiched between two metallic leads is always described by a robust Fermi liquid fixed point, and displays perfect dc conductance.}

While this result may sound counterintuitive, it arises in a natural way from the normal-state proximity effect. Consider a multilayered device composed of metallic leads sandwiching a finite number of Mott-insulating `barrier' planes. At sufficiently high temperatures, the Mott insulator develops a thermal excitation induced  (one electron) density of states (DOS) within the Mott gap, and the system conducts electricity.  As the temperature is lowered, this DOS decreases, but can never become smaller than the DOS induced via tunneling due to the normal-state proximity effect from the metallic leads.  The tunneling DOS at zero frequency is expected to decrease exponentially with thickness as one moves deeper into the barrier planes. But in no circumstance does it vanish.  Hence, as the temperature is lowered, in a pure system, a low temperature Fermi liquid must form. Eventually, at $T=0$, the electronic self-energy vanishes at $\omega=0$, and the system restores full periodicity, with a well defined Fermi surface, and no scattering.  {\em The device develops a perfect conducting channel.}  The temperature scale that governs this behavior is the effective Fermi temperature of the innermost barrier plane; hence this state is very fragile with respect to increasing $T$, to the passage of finite frequency currents, and to the presence of a voltage across the planes,  disorder, or magnetic order in the system (we assume a paramagnetic phase in this work). However, for thin enough barriers, where the low-energy scale is not exponentially small, one can create unique devices with highly nonlinear properties as a consequence. For example, the system will pass dc current up to some maximal current with ease, but then rapidly make a transition to a tunneling response as the current gets too large (a current limiter); and similarly if the driving fields change with too rapid a frequency, or if the system is in the presence of a high enough temperature or applied voltage.  While this result can become one of merely academic interest, not relevant to experimental systems, if the temperature scales are too low, our numerical modeling shows that {\em there is a wide parameter range where one should be able to see this phenomena}. Note that the conclusion of our work, that a Mott insulator will always develop a robust low-temperature conducting channel, is {\em contrary} to the conclusion reached in recent work on normal-metal--Mott-insulating interfaces~\cite{helmes08}, where the authors state: ``our results imply that the Mott insulator is {\it de facto} impenetrable to the metal''.
%of that work improperly confused their finite temperature calculations with the behavior that will occur when one is at $T=0$, and thereby did not notice that the Fermi liquid behavior they observed naturally leads to the opening of a conducting channel at $T=0$.

We take the barrier of our metal-barrier-metal sandwich to be described by the single-band Hubbard model, and the metallic leads by the noninteracting tight-binding model.  The Hamiltonian for the system is
\begin{eqnarray}
\mathcal{H}&=&-\sum_{ij\alpha\sigma}t_{ij}^\parallel c^\dagger_{i\alpha\sigma}c^{}_{j\alpha\sigma}- t \sum_{i\alpha\sigma} [c^\dagger_{i\alpha\sigma}c^{}_{i\alpha+1\sigma}+h.c.]\nonumber\\
&-&\mu\sum_{i\alpha\sigma}c^\dagger_{i\alpha\sigma}c^{}_{i\alpha\sigma}+\sum_{i\alpha}U_\alpha(n_{i\alpha\uparrow} -\frac{1}{2})(n_{i\alpha\downarrow}-\frac{1}{2}).\;
\end{eqnarray}
Here the label $\alpha$ indexes the planes, and the label $i$ indexes sites of the two-dimensional square lattice in each plane. The operator $c^\dagger_{i\alpha\sigma}$ ($c^{}_{i\alpha\sigma}$) creates (destroys) an electron of spin $\sigma$ at site $i$ on the plane $\alpha$. We set the in-plane hopping $t^{\parallel}$ to be nearest neighbor only, and equal to $t$, the hopping between planes, so that the lattice structure is that of a simple cubic lattice.  The interaction term is written in a particle-hole symmetric fashion, with $U_\alpha = U$ for the barrier planes, and zero for the metallic planes. Hence the chemical potential $\mu$ is equal to zero for half filling,  both for the half-filled ballistic metal and the half-filled Hubbard system. The system, assumed to be at half filling,  is then particle-hole symmetric; so we drop $\mu$ hereafter, and also set $t = 1$. Furthermore, we assume a paramagnetic regime, and drop all spin indices in the formulas below.

The many-body problem is solved within a self-consistent inhomogeneous dynamical mean-field theory (IDMFT) framework, using the method of Potthoff and Nolting~\cite{potthoff99,freericks_idmft,freericks_book}. That is, we assume that the self-energy is local and the same for all the sites in a barrier plane, but can vary from plane to plane, i.e., depends on the index $\alpha$.
%We take 30 self-consistent metallic planes on each side of the barrier with a thickness of $N$ barrier planes. The boundary conditions are set by demanding that deep
%inside the metallic leads, physical properties approach that of the bulk metal. Hence, we model a total of $N + 60$ planes number the planes from $-29$ to $N+30$, so that the barrier planes are numbered from $1$ to $N$. Then  the equalities
%$G_{N+30} \approx G_{\infty} = G_{bulk}$ and $G_{-29} \approx G_{-\infty} = G_{bulk}$ hold.  The standard IDMFT algorithm for solving the properties of this system has appeared elsewhere~\cite{freericks_idmft,freericks_book}, and we do not repeat the formalism here.
We label the barrier planes with $\alpha$ values from 1 to N. Thus $\alpha \geq (N+1)$ or $\alpha \leq 0$ correspond to the metallic layers.
The local Green's function at plane $\alpha$ is found by employing the quantum zipper algorithm, which expresses the Green's function in terms of two continued fractions, one from the left, $L_\alpha(\epsilon,\omega)$ and one from the right, $R_\alpha(\epsilon,\omega)$ as
\begin{equation}
 G_\alpha(\omega)=\int d\epsilon \rho_{2d}(\epsilon)\frac{1}{L_\alpha(\epsilon,\omega)+R_\alpha(\epsilon,\omega)-Z_\alpha(\omega)+\epsilon},
\label{eq: g_local}
\end{equation}
where $\rho_{2d}(\epsilon)$ is the noninteracting DOS for the square-lattice and $Z_\alpha(\omega) \equiv \omega - \Sigma_\alpha(\omega)$ , with $\Sigma_\alpha(\omega)$, the local self-energy for plane $\alpha$, being non-zero only for $\alpha = 1,...,N$, i.e., the barrier planes. The left and right functions are given by
%$L_\alpha(\epsilon,\omega) = Z_\alpha(\omega) - \epsilon - [L_{\alpha-1}(\epsilon,\omega)]^{-1}$ and %$R_\alpha(\epsilon,\omega) = Z_\alpha(\omega) - \epsilon - [R_{\alpha+1}(\epsilon,\omega)]^{-1}$ respectively, with
\begin{widetext}
\begin{equation}
  L_\alpha(\epsilon,\omega)=Z_\alpha(\omega)-\epsilon-\frac{1}{Z_{\alpha-1}(\omega)-\epsilon- {\displaystyle \frac{1}{Z_{\alpha-2}(\omega)-\epsilon-
  {\displaystyle \mathop{\,}_{~~\ddots~
 {\mathop{\,}_{\displaystyle{ {\displaystyle ~ Z_{1}-\epsilon- {\displaystyle \frac{1}{\frac{1}{2}(\omega - \epsilon)\pm\frac{1}{2}\sqrt{(\omega -
\epsilon)^2-4}}}} }}}}}}}}
\label{eq:L-eqn}
\end{equation}
and
\begin{equation}
 R_\alpha(\epsilon,\omega)=Z_\alpha(\omega)-\epsilon-\frac{1}{Z_{\alpha+1}(\omega)-\epsilon- {\displaystyle \frac{1}{Z_{\alpha+2}(\omega)-\epsilon-
   {\displaystyle \mathop{\,}_{~~\ddots~
   {\mathop{\,}_{\displaystyle{{\displaystyle ~ Z_{N}-\epsilon- {\displaystyle \frac{1}{\frac{1}{2}(\omega - \epsilon)\pm\frac{1}{2}\sqrt{(\omega -
\epsilon)^2-4}}}}}} }}}}}}
\label{eq:R-eqn}
\end{equation}
\end{widetext}
Furthermore, $ L_\alpha(\epsilon,\omega) = g^{-1}_{1d} (\omega-\epsilon) = R_{\alpha^\prime}(\epsilon,\omega) $ for $\alpha \leq 0$ ($L_\alpha$) and $\alpha^\prime \geq (N+1)$ ($R_{\alpha^\prime}$), where $ g^{-1}_{1d} (\omega) \equiv \frac{1}{2}\omega  \pm \frac{1}{2}\sqrt{\omega^2-4} $ is the inverse local Green's function for a 1-d tight-binding band. The $\pm$ sign in the above equations has to be chosen so that in the upper half of the complex $\omega$ plane $L_\alpha$, $R_{\alpha}$ and $g^{-1}_{1d}$ are analytic, have a positive imaginary part and go asymptotically as $\omega$ for large $|\omega|$.
Thus the IDMFT formalism leads to N separate impurity problems, one for each barrier plane, which need to be solved to obtain $\Sigma_\alpha(\omega)$. These are coupled via Eqs.~(\ref{eq: g_local}--\ref{eq:R-eqn}) to determine  $G_\alpha(\omega)$, and hence the $N$ bath Green's functions. The details of the IDMFT algorithm are well described elsewhere\cite{freericks_idmft,freericks_book}, and we do not repeat them here. Note that in the present context $G_\alpha(\omega)$ for the metallic planes do not enter the IDMFT recursions, and can be calculated at the end for any $\alpha$. Within the IDMFT framework the above procedure therefore fully solves the problem of a finite number of barrier layers sandwiched between {\em semi-infinite metallic leads}.

On the basis of these equations we argue below that the DOS for each plane in the barrier is bounded from below by the tunneling result, hence the normalized hybridization parameter $\Gamma_{o\alpha}$ for the impurity problem at plane $\alpha$ is nonzero for all $\alpha$, and {\em cannot iterate to zero} via the self-consistent IDMFT algorithm.  This then implies that there is an effective Fermi temperature $T_{F\alpha}\propto\exp[-8U/\pi\Gamma_{0\alpha}]$ for each of the Mott-insulating planes, and when $T$ is lowered below $T_{F\alpha}$, that plane will be in a Fermi-liquid state, leading to a vanishing self-energy at $\omega=0$ and $T=0$.  For a thick barrier, with a large $U$, these temperature scales are exponentials of exponentially large negative numbers, and are hence exceedingly small, and would never be observed in the laboratory.  But for thin barriers, with $U$ close to the critical $U$ for the Mott transition, the effective Fermi temperatures are much higher, and could become accessible experimentally.

%\begin{equation}
% L_\alpha(\epsilon,\omega)=Z_\alpha(\omega)-\epsilon-\cfrac{1}{Z_{\alpha-1}(\omega)-\epsilon-\cfrac{1}{Z_{\alpha-2}(\omega)-\epsilon-\dots -\cfrac{1}{Z_{1}-\epsilon-\cfrac{1}{\frac{1}{2}(\omega - \epsilon)\pm\cfrac{1}{2}\sqrt{(\omega - \epsilon)^2-4}}}}}
%\end{equation}
%%\begin{equation}
%%  L_\alpha(\epsilon,\omega)=Z_\alpha(\omega)-\epsilon-\frac{1}{Z_{\alpha-1}(\omega)-\epsilon- {\displaystyle \frac{1}{Z_{\alpha-2}(\omega)-\epsilon-
%%  {\displaystyle \mathop{\,}_{~~\ddots~
%% {\mathop{\,}_{\displaystyle{ {\displaystyle ~ Z_{1}-\epsilon- {\displaystyle \frac{1}{\frac{1}{2}(\omega - \epsilon)\pm\frac{1}{2}\sqrt{(\omega -
%%\epsilon)2-4}}}} }}}}}}}}
%\end{equation}
%and
%%\begin{equation}
%% R_\alpha(\epsilon,\omega)=Z_\alpha(\omega)-\epsilon-\cfrac{1}{Z_{\alpha+1}(\omega)-\epsilon-\cfrac{1}{Z_{\alpha+2}(\omega)-\epsilon-\cdots -\cfrac{1}{Z_{N}-\epsilon-\cfrac{1}{\frac{1}{2}(\omega - \epsilon)\pm\cfrac{1}{2}\sqrt{(\omega - \epsilon)^2-4}}}}}.
%%\end{equation}
%\begin{equation}
% R_\alpha(\epsilon,\omega)=Z_\alpha(\omega)-\epsilon-\frac{1}{Z_{\alpha+1}(\omega)-\epsilon- {\displaystyle \frac{1}{Z_{\alpha+2}(\omega)-\epsilon-
%   {\displaystyle \mathop{\,}_{~~\ddots~
%   {\mathop{\,}_{\displaystyle{{\displaystyle ~ Z_{N}-\epsilon- {\displaystyle \frac{1}{\frac{1}{2}(\omega - \epsilon)\pm\frac{1}{2}\sqrt{(\omega -
%\epsilon)2-4}}}}}} }}}}}}
%\end{equation}
%\end{widetext}

The tunneling DOS, being determined by the proximity effect, gets smaller, and hence the self-energy larger, as we move deeper into the barrier. We therefore assume that
\begin{eqnarray}
 |\Sigma_{N/2}(\omega\approx 0)|&\gg& |\Sigma_{N/2\pm 1}(\omega\approx 0)|\gg |\Sigma_{N/2\pm 2}(\omega\approx 0)|\nonumber\\
&\gg&\cdots\gg|\Sigma_{N/2\pm N/2}(\omega\approx 0)|.
\end{eqnarray}
Hence we can write
\begin{equation}
 L_\alpha(\epsilon,\omega)\approx Z_\alpha(\omega)-\epsilon+\bar\gamma^L_{\alpha}(\epsilon,\omega)
\label{eq: l_approx}
\end{equation}
and
\begin{equation}
 R_\alpha(\epsilon,\omega)\approx Z_\alpha(\omega)-\epsilon+\bar\gamma^R_{\alpha}(\epsilon,\omega).
\label{eq: r_approx}
\end{equation}
Here $\bar\gamma^L_{\alpha}(\epsilon,\omega)=1/\Sigma_{\alpha-1}(\omega)$ for $\alpha>1$, and %$\bar\gamma^L_1(\epsilon,\omega)=-(\omega - \epsilon)/2\pm\sqrt{(\omega - \epsilon)^2-4}/2$ for $\alpha=1$,
$\bar\gamma^L_1(\epsilon,\omega)= -g_{1d}(\omega~-~\epsilon)$ for $\alpha=1$. Similarly,
$\bar\gamma^R_{\alpha}(\epsilon,\omega)=1/\Sigma_{\alpha+1}(\omega)$ for $\alpha<N$, %$\bar\gamma^R_1(\epsilon,\omega)=-(\omega - \epsilon)/2\pm\sqrt{(\omega - \epsilon)^2-4}/2$ for $\alpha=N$.
and $\bar\gamma^R_N(\epsilon,\omega) = -g_{1d}(\omega-\epsilon)$ for $\alpha=N$.
Substituting into Eq.~(\ref{eq: g_local}), using the result $|\Sigma_\alpha(\omega\approx 0)|\gg |\bar\gamma_\alpha^{L,R}(\epsilon,\omega)|$, expanding the asymptotic form of the fraction and performing the integration over $\epsilon$ then yields
\begin{equation}
 G_\alpha(\omega\approx0)\approx -\frac{1}{\Sigma_\alpha(\omega\approx 0)}\left [ 1+\frac{\omega + \gamma_\alpha^L+\gamma_\alpha^R}{\Sigma_\alpha(\omega\approx 0)}+\cdots \right ],\;
\end{equation}
with $\gamma_\alpha^{L,R}=\int d\epsilon \rho_{2d}(\epsilon)\bar\gamma_\alpha^{L,R}(\epsilon,\omega\approx 0)$.  Performing the integrals yields $\gamma_\alpha^L=1/\Sigma_{\alpha-1}(\omega\approx 0)$ for $\alpha>1$, $\gamma_1^L=0.525i$, $\gamma_\alpha^R=1/\Sigma_{\alpha+1}(\omega\approx 0)$ for $\alpha<N$, and $\gamma_N^R=0.525i$. Using this result, we immediately find that
\begin{equation}
 \Gamma_{0\alpha}={\rm Im} \, G_{0\alpha}^{-1}(\omega\approx 0)={\rm Im}(\gamma_\alpha^L+\gamma_\alpha^R),
\label{eq: gamma_0}
\end{equation}
with $G_{0\alpha}$ being the effective medium for the impurity, which satisfies $G_{0\alpha}^{-1}=G_\alpha^{-1}+\Sigma_\alpha$.
In general, this result is a {\it lower bound} for $\Gamma_{0\alpha}$, as the DOS may never achieve this value as $T$ is lowered from high temperature; this implies that our estimates for $T_{F\alpha}$ will also be lower bounds.

The {\em bulk} insulator phase of the Hubbard model at half filling is characterized by a pole in the self-energy that takes the form
\begin{equation}
 \Sigma_{\rm bulk}(\omega\approx 0)=\frac{r(U)}{\omega+i\delta},
\end{equation}
where $r(U)$, the residue of the pole, vanishes when the insulating state is no longer stable, i.e., when $U< U_{c1}\approx 11.4$ for the simple cubic lattice solved with DMFT (the metallic phase disappears for $U>U_{c2}\approx 13.1$). For the heterostructure, it is interesting to explore the consequences of the ansatz
\begin{equation}
 \Sigma_{\alpha}(\omega\approx 0)=\frac{r(U)}{\omega+\gamma_\alpha^L+\gamma_\alpha^R},
\end{equation}
because this is the form that arises in an exact solution of the impurity problem for the Falicov-Kimball model in the insulating phase, and in the Hubbard III approximation for the Hubbard model with the $\Gamma_{0\alpha}$ in Eq.~(\ref{eq: gamma_0}). In this case, the imaginary part of the denominator never vanishes, and instead is forced to be nonzero due to the normal-state proximity effect.  Using this result and the definitions of the $\gamma$ coefficients, we obtain, for odd $N$,
\begin{equation}
% G_\alpha(\omega\approx 0)=\left [ \begin{matrix}
%                                    -\frac{0.525i}{[r(U)]^\alpha} & 1\le\alpha\le \frac{N-1}{2}\\
%                                    -\frac{1.05i}{[r(U)]^{\frac{N+1}{2}}} & \alpha=\frac{N+1}{2}\\
%                                    -\frac{0.525i}{[r(U)]^{N+1-\alpha}} & \frac{N+3}{2}\le \alpha\le N
%                                   \end{matrix}\right .
\Gamma_{0\alpha}(\omega\approx 0) \gtrsim \left [ \begin{matrix}
                                    0.525 / [r(U)]^{\alpha - 1} & 1\le\alpha\le \frac{N-1}{2}\\
                                    1.05  / [r(U)]^{\frac{N-1}{2}} & \alpha=\frac{N+1}{2}\\
                                    0.525 / [r(U)]^{N-\alpha} & \frac{N+3}{2}\le \alpha\le N
                                   \end{matrix}\right .
\end{equation}
%in addition, we have $\Sigma_\alpha(\omega\approx 0)\approx -1/G_\alpha(\omega\approx 0)$ and   $\Gamma_{0\alpha}\approx -{\rm Im} G_{\alpha}(\omega\approx 0)r(U)$. 
Note that this behavior corresponds to an exponential decay of the tunneling DOS at $\omega=0$, with a correlation length given by $1/\ln r(U)$. In the Falicov-Kimball model on a simple cubic lattice, one has $r(U)=(U^2-24)/4$~\cite{demchenko}; for $U=6$, just slightly above the critical $U$ for the Mott transition at $U_c=2\sqrt{6}$, the correlation length is $0.62$~\cite{freericks_book} and the analytic approach predicts it to be $0.92$.  For the Hubbard model on a simple cubic lattice, a {\em fit obtained from numerics} (see below) gives $r(U) \simeq (U^2-U^2_{c1})/4 -~0.13(U-U_{c1}) +~13.4\sqrt{U-U_{c1}} -~52.3\ln{(U/U_{c1})} $.  This suggests that at $T=0$ each barrier plane of the metal--Mott-insulator--metal structure will become a Fermi liquid, leading thereby to a half-filled, simple cubic lattice Fermi surface and perfect (ballistic) conductivity. (The conductance will be finite, because the ballistic conductance of the simple cubic lattice is found by counting the number of conducting channels [via the Fermi surface area] and multiplying by the quantum of conductance -- this will look like a ``contact conductance''.)

\begin{figure}[htb]
\centering
\includegraphics[width=3.0in,clip=on]{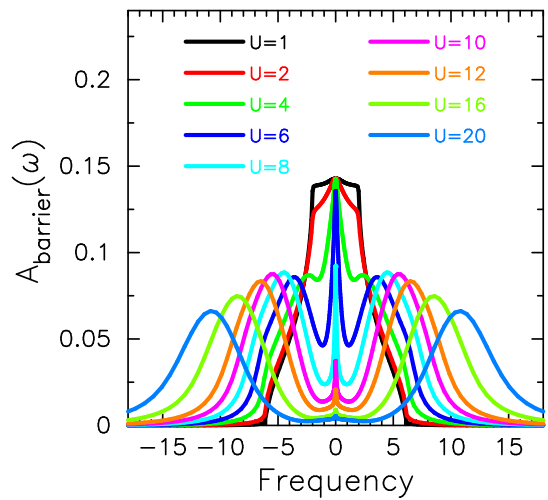}
\caption{(Color online.) Local spectral function at the barrier plane for a single-barrier-plane multilayer, for $T=0.01$, and a range of $U$'s (increasing from top to bottom near zero frequency).}
\label{fig: n=1}
\end{figure}

Next, as a complement to the above arguments, we present our numerical work where we solve the impurity problem on each interacting plane with the numerical renormalization group (NRG)~\cite{nrg_review}. We
choose the discretization parameter $\Lambda=2$ and we keep $1200$ states at each NRG iteration, and calculate the local DOS or one particle spectral function on different planes, given by $A_\alpha(\omega)=-{\rm Im}G_\alpha(\omega)/\pi$.

Fig.~\ref{fig: n=1}  shows our results for the DOS at the barrier plane for the single-plane barrier ($N=1$) multilayer, for $T=0.01$ and different values of $U$.   Note how one can clearly see the Fermi liquid coherence peak forming for all values of $U$, but the coherence is not complete for the larger $U$ values because the effective Fermi temperature of the central plane falls below our chosen low temperature of $T=0.01$ for those $U$ values; in addition, note how the Fermi peak is narrower for larger $U$ values for the same reason.

\begin{figure}[htb]
\centering
\includegraphics[width=3.0in,clip=on]{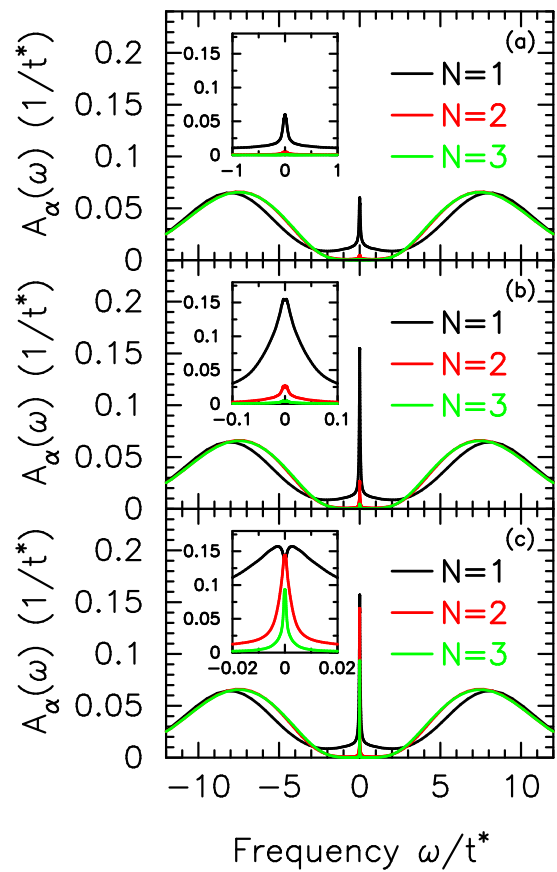}
\caption{(Color online.) Local spectral function of the five-barrier-plane multilayer for $U = 16$, and three different temperatures: (a) $T=0.01$, (b) $T=0.001$, and (c) $T=0.0001$. We show the three inequivalent barrier planes ($N=1$ is next to the interface, $N=3$ is at the center).  The insets show blow-ups of the low frequency region.}
\label{fig: n=5}
\end{figure}

Fig.~\ref{fig: n=5} shows our results for a five-plane ($N=5$) device at three different temperatures for $U = 16$. We see that as the temperature is lowered the spectral functions on all the layers tend to the noninteracting simple cubic DOS value at $\omega=0$, but the temperature at which the Fermi peak develops is lower for the deeper planes, and the width is significantly narrower.  We estimate the Fermi temperatures for the different planes to be $T_{F1} \approx 0.015$, $T_{F2} \approx 0.0009$ and $T_{F3} \approx 0.00025$. The high energy features in the DOS remain unchanged as $T$ is lowered.

In so far as transport is concerned, our numerical techniques are not yet reliable enough to accurately determine the conductance of these devices with a Kubo formula.  Nevertheless, we conjecture that decreasing the temperature is tantamount to decreasing the length of the barrier as more and more insulating layers
close to the interface become metallic. We speculate that this would result in a strong nonlinear dependence of the dc conductance on temperature, which will be tunneling like, but with sharp increases as T is lowered below the effective Fermi temperature of each symmetric pair of planes.  Finally, when T is lowered below the Fermi temperature of the central planes, the system will become a perfect dc conductor. This is clearly the case in the infinite-dimensional limit where vertex corrections vanish. We also expect a very narrow Drude
peak at low temperatures, which should be observable using optical or other probes of ac conductivity.  However, disorder, which is always present in
any real device, could in practice suppress this low temperature behavior so that the metallic regime  may be hard to see in real samples, especially in those with large Mott-Hubbard gaps, magnetic order, or thick insulating barriers.

In summary, the main result of this work is that {\em a system where a Mott insulating barrier is sandwiched between metallic leads becomes a ballistic metal at sufficiently low temperatures no matter how strong the interaction $U$, as long as the barrier thickness is finite}. This striking result is similar to the zero-bias anomaly of the quantum dot problem in the Kondo regime~\cite{Wolff}. The particle-hole symmetric system (with the same hopping for all planes and directions) we are considering here has the special feature that at zero temperature and zero frequency it becomes translational invariant and equivalent to the noninteracting problem.  We also conjecture this Fermi-liquid behavior will hold in cold-atom systems trapped in a harmonic trap on an optical lattice.  The central Mott region will be surrounded by normal metal, which should create a proximity effect via tunneling, and the entire system will become a Fermi liquid as $T\rightarrow 0$, so the Mott phase will disappear at zero frequency.  It is not clear such low temperatures could ever be achieved in the laboratory.

{\it Acknowledgments}. H. Z. and J. K. F. acknowledge support from the National Science Foundation under grant number DMR-0705266.
H. R. K. is supported under ARO Grant W911NF0710576 with funds from the DARPA OLE Program, and from the Department of Science and Technology, India as a J. C. Bose National Fellow.
Th. P. acknowledges support from the collaborative research center (SFB) 602.
%\bibliography{../../Bibliography/mybib}

\begin{thebibliography}{9}

\bibitem{hwang}
A. Ohtomo and  H. Y. Hwang, Nature {\bf 427}, 423 (2004).

\bibitem{thiel06} S. Thiel, et al., Science \textbf{313}, 1942 (2006).

\bibitem{helmes08} R. W. Helmes, T. A. Costi, and A. Rosch, Phys. Rev. Lett. {\bf 101}, 066802 (2008).

\bibitem{potthoff99} M. Potthoff and W. Nolting, Phys. Rev. B \textbf{59}, 2549 (1999).

\bibitem{freericks_idmft}
J. K. Freericks, Phys. Rev. B {\bf 70}, 195342 (2004).

\bibitem{freericks_book} J. K. Freericks, \emph{Transport in multilayered nanostructures: the dynamical mean-field theory approach} (Imperial College Press, London, 2006).

\bibitem{demchenko}
D. O. Demchencko, A. V. Joura, and  J. K. Freericks,  Phys. Rev. Lett. {\bf 92}, 216401 (2004).

\bibitem{nrg_review}
R. Bulla, T. A. Costi and Th. Pruschke, Rev. Mod. Phys. {\bf 80}, 395 (2008), and references therein.

\bibitem{Wolff} The one dimensional analog of the $N=1$ case considered above, corresponding to a tight-binding chain with a non-zero $U$ only at one cental site, is equivalent to the Wolff or local spin fluctuation model. As discussed in H. R. Krishnamurthy, K. G. Wilson and J. W. Wilkins, Phys. Rev. B {\bf 21}, 1003 (1980), it belongs to the same universality class as the Anderson impurity problem. At $T=0$ it has a Kondo resonance at low frequencies, corresponding to a local Fermi liquid, no matter how large U is, which again translates into ballistic dc conductance.

%\bibitem{bulla01} R. Bulla, T. A. Costi, and D. Vollhardt, Phys. Rev. B \textbf{64}, 045103 (2001).

\end{thebibliography}

\end{document}